\documentclass[12pt]{iopart}

\newif\ifshowhlRone \showhlRonefalse
\newif\ifshowhlRtwo \showhlRtwofalse
\newif\ifshowhlRthree \showhlRthreefalse
\newif\ifshowall \showalltrue
\newif\ifpreprint \preprinttrue
\newif\ifdarkmode \darkmodefalse

\usepackage{add-ons}
\begin{document}

\title{Revisiting the apparent horizon finding problem with multigrid methods}

\author{Hon-Ka Hui\footnote[1]{Author to whom any correspondence should be addressed.}\,\orcidlink{0000-0002-2815-527X}, Lap-Ming Lin\,\orcidlink{0000-0002-4638-5044}}

\address{Department of Physics, The Chinese University of Hong Kong, Hong Kong SAR, China}
\ead{hkhui@link.cuhk.edu.hk, lmlin@cuhk.edu.hk}
\vspace{10pt}

\begin{abstract}
    Apparent horizon plays an important role in numerical relativity as it provides a 
    tool to characterize the existence and properties of black holes on three-dimensional spatial slices in 3+1 numerical spacetimes. Apparent horizon finders based on different techniques have been developed. 
    In this paper, we revisit the apparent horizon finding problem in numerical relativity using multigrid-based algorithms. 
 We formulate the nonlinear elliptic apparent horizon equation as a linear Poisson-type equation with a nonlinear source, and solve it using a multigrid algorithm with Gauss-Seidel line relaxation.
 A fourth order compact finite difference scheme in spherical coordinates is derived and employed to reduce the complexity of the line relaxation operator to a tri-diagonal matrix inversion. 
    The multigrid-based apparent horizon finder developed in this work is capable of locating apparent horizons in generic spatial hypersurfaces without any symmetries.
    The finder is tested with both analytic data, such as Brill-Lindquist multiple black hole data, and numerical data, including off-centered Kerr-Schild data and dynamical inspiraling binary black hole data. 
    The obtained results are compared with those generated by the current fastest finder \textsc{AHFinderDirect} (Thornburg, \textit{Class.~Quantum Grav.}~\textbf{21}, 743, 2003), which is the default finder in the open source code \textsc{Einstein Toolkit}. 
Our finder performs comparatively in terms of accuracy, and starts to outperform \textsc{AHFinderDirect}
at high angular resolutions ($\sim 1^\circ$) in terms of speed. Our finder is also more flexible to initial guess, as opposed to the Newton's method used in \textsc{AHFinderDirect}. This suggests that the multigrid algorithm provides an alternative option for studying apparent horizons, especially when high resolutions are needed. 
\end{abstract}


\section{Introduction}
\label{sec:1-intro}
    Apparent horizon has been an important concept for the numerical studies of spacetime that involve black holes. It serves as a real-time indicator of the existence of a black hole in numerical simulations, as its presence implies the presence of a surrounding event horizon \cite{Hawking_Ellis_1973}. The determination of apparent horizon during numerical simulations is made possible because \hlRone{it can be defined locally in time}. This is in contrast to the teleological nature of event horizon, whose determination requires the full knowledge of the future evolution of a spacetime. Knowing the presence of an apparent horizon during simulation is important not only because it allows the estimation of the properties of the black hole, such as its mass and spin, but it also marks the region within which singularity may appear at a later time \cite{Penrose1965}, such region sometimes is necessary to be avoided through excision to maintain stability of the simulation \cite{Sykes2023} (also see, e.g., \cite{Alcubierre2003, Baiotti2006} for using singularity-avoiding gauge instead of excision).

    The dynamical evolution of the apparent horizon is also of interest for understanding black holes and gravitational wave physics. The notion of dynamical horizons (e.g., \cite{Ashtekar2003, Schnetter2006}, also see \cite{Ashtekar2004_review} for a review), which is loosely speaking 
    a world tube formed by the foliation of the apparent horizon at different time slices, can be used instead of event horizons to better study the interaction and merging of black holes \cite{Pook-Kolb2021}. Recent studies have also suggested correlation between the dynamics of (common) apparent horizon in black hole binaries with the gravitational waves emitted \cite{Evans2020, Prasad2020}, which could hint a possible imprint of event horizon properties on the signal.

    Numerous efforts have been dedicated to developing an efficient apparent horizon finder, as discussed in a comprehensive review \cite{Thornburg2007}. 
    Currently, the fastest algorithm for locating the apparent horizon is based on Newton's method (\cite{Thornburg2003, Schnetter2003}), along with some variants of the matrix inversion scheme for the associated Jacobi equation. In general, such a method suffers from the poor scaling of the computational time complexity and the reliance on a good initial guess. On the other hand, while multigrid methods in general exhibit much better scaling, they have not been applied to the apparent horizon searching problem to our best knowledge. Multigrid methods, owing to their efficiency in solving elliptic partial differential equations (PDEs), have been applied to different problems in numerical relativity, including the initial data construction of binary stellar objects (see, e.g., \cite{Baumgarte1998, Brandt1997, Cook1993, East2012, Moldenhauer2014}) and more recently to a relativistic hydrodynamic code under the conformally flat assumption \cite{Cheong2020}. It should also be pointed out that a review article introducing the multigrid method for numerical relativists was already written about 40 years ago \cite{Choptuik1986}. With the expectation that multigrid methods can solve elliptic equations efficiently, we aim to implement an apparent horizon finder in a multigrid approach and examine whether this could be a better algorithm in terms of speed and robustness, especially for cases when one needs to resolve the apparent horizon with high enough resolution to increase the precision of the inferred properties of the black hole in simulations. 
    
    This paper is organized as follows. In Section \ref{sec:2-method}, we first describe the equation to solve for the apparent horizon, and present the implementation of the multigrid algorithm in our apparent horizon finder. Section \ref{sec:3-tests} includes the test results of our finder in terms of accuracy and efficiency in several benchmark data. Finally, we summarize our work and give future prospects in Section \ref{sec:4-conclusion}.

\section{Method and Implementation}
\label{sec:2-method}
    Locating the apparent horizon requires solving a non-linear elliptic PDE---the expansion equation. In this section, we first introduce the variant ansatz of the expansion equation we employed. We then give a brief overview of the multigrid methods in general and the specific scheme we used in this work.

    \subsection{Apparent Horizon and the Expansion Equation}
        \hlRone{A black hole region is characterized by the event horizon, which is defined as the boundary 
        of the causal past of the future null infinity~{\cite{Hawking_Ellis_1973}}.}
        \hlRtwo{%
          The event horizon is therefore a global property of the underlying 4-dimensional spacetime, and its determination in a numerical simulation can only be done approximately by integrating null geodesics (e.g., \mbox{\cite{Hughes_1994}}) or null surfaces (e.g., \mbox{\cite{Anninos_1995}}) backwards in time once the spacetime has essentially settled down to a final stationary state (see \mbox{\cite{Thornburg2007}} for the review of these algorithms).}
        \hlRone{%
          However, if one would like to trace the motion of a black hole during a simulation, it is important 
          to have a local (in time) characterization of the black hole. This can be achieved by the concept of apparent horizon.}

        \hlRone{%
          Intuitively, a black hole is a region where light rays are ``trapped'' and cannot escape to infinity. This idea can be formalized by introducing the notion of a trapped region in a 3-dimensional spatial slice embedded in the 4-dimensional spacetime, defined as the union of all trapped surfaces\footnote{%
            \hlRone{A trapped surface is a closed spacelike 2-surface embedded in the 3-dimensional 
            spatial slice with the property that the expansion function $\Theta$ of both future-pointing outgoing and ingoing null geodesics is negative}
            \hlRtwo{everywhere on the surface}.%
            }. 
          The outer boundary of the trapped region is defined to be the apparent horizon. However, this mathematical definition of the apparent horizon is not 
          convenient for practical use in numerical simulations. In the following, we shall provide a slightly different definition of the apparent horizon that is used in numerical relativity, using the expansion function $\Theta$. We also describe how 
          $\Theta$ depends on the spacetime geometric variables, and the ansatz we adopted to solve for an apparent horizon.
        }

        \subsubsection{Notations and definitions}
            Let $n^a$ denote the timelike future-pointing unit normal vector to a spacelike hypersurface $\Sigma_t$, and $s^a$ denote the spacelike outward-pointing unit normal vector to an arbitrary 2-dimensional smooth closed surface $\mathcal{S}$ embedded in $\Sigma_t$. The spatial 3-metric $\gamma_{ab}$ induced on $\Sigma_t$ and the 2-metric $m_{ab}$ induced on $\mathcal{S}$ are given by
            \begin{numparts}
            \begin{eqnarray}
                \gamma_{ab} &= g_{ab} + n_a n_b,
                \\
                m_{ab} &= \gamma_{ab} - s_a s_b = g_{ab} + n_a n_b - s_a s_b ,
            \end{eqnarray}
            \end{numparts}%
            where $g_{ab}$ is the spacetime metric.
            By denoting the unit vector along the future-pointing outgoing null geodesics by $k^a = \left( s^a + n^a \right)/\sqrt{2}$, the expansion function can be expressed as (see, e.g., \cite{Baumgarte_Shapiro_2010}),
            \begin{equation}
            \label{eq:expansion_function}
                \Theta \equiv m^{ab} \nabla_a k_b = (D_i s^i + K_{ij} s^i s^j - K )/\sqrt{2},
            \end{equation}
            where $\nabla_a$ and $D_a$ are the covariant derivatives associated with the metric tensors $g_{ab}$ and $\gamma_{ab}$, respectively; $K_{ij}$ is the extrinsic curvature, and $K$ denotes its 
            trace (i.e., $K \equiv \gamma^{ij} K_{ij}$).
            \ifshowhlRtwo\hlRtwo{%
              \mbox{\sout{(\ldots)}}}\fi
            \hlRone{%
            $\mathcal{S}$ is called a 
            marginally outer trapped surface (MOTS) if $\Theta=0$ everywhere on $\mathcal{S}$. There could also be multiple MOTSs or even nested MOTSs in $\Sigma_t$.} 
            The apparent horizon is defined to be the outermost of such surfaces.
            
            By assuming that the apparent horizon is topologically equivalent to a 2-sphere and is a star-shaped surface around an interior local coordinate origin\footnote{%
                We refer the reader to \cite{Pook-Kolb2019} for the removal of the star-shaped assumption in the axisymmetric case.}, 
            we use standard spherical coordinates $(r, \theta, \phi)$ to parameterize it. The apparent horizon surface is represented by a horizon function $h(\theta, \phi)$ which measures the radial coordinate distance of the surface from the local origin. The aforementioned outward-pointing normal $s^i$ can be constructed using the gradient of a level-set function $F= r- h(\theta, \phi)$, namely
            \begin{equation}
                s_i = \lambda m_i \equiv \lambda \partial_i F = \lambda (1, -\partial_\theta h, -\partial_\phi h),
            \end{equation}
            where $\lambda$ is the normalization factor such that $s_i s^i = 1$. Using such 
            parameterization, (\ref{eq:expansion_function}) can be cast into an elliptic PDE in terms of $h$.

        \subsubsection{The {\it{ansatz}} adopted}
            Following \cite{Lin2007}, we separate out a linear elliptic operator from the non-linear expansion equation $\Theta=0$, such that it is in a suitable form to be solved by standard relaxation method. We first consider
            \begin{equation}
            \label{eq:ah_master_propose}
                \Delta_{\theta\phi} h - 2 h = \Delta_{\theta\phi} h - 2 h + \rho \Theta.
            \end{equation}
            where $\rho = \rho(h, \theta, \phi)$ is a scalar function to be determined and $\Delta_{\theta\phi}$ is the flat-space Laplacian on a 2-sphere which is given by
            \begin{equation}
                \Delta_{\theta\phi} h = \partial^2_\theta h + \cot\theta \,\partial_\theta h + \frac{1}{\sin^2\theta} \partial^2_\phi h.
            \end{equation}
            We briefly discuss how the Laplacian shows up in the expansion equation and how the scalar function $\rho$ should be chosen.
            
            A 3-metric tensor $\gamma_{ij} (r, \theta, \phi)$ on a spacelike hypersurface $\Sigma_t$ can be expressed by its conformally related metric $\bar{\gamma}_{ij}$ and be further decomposed into the flat metric $\eta_{ij}$ along with a tensor field $h_{ij}$, i.e.,
            \begin{equation}
                \gamma^{ij} = \psi^{-4} \bar{\gamma}^{ij} = \psi^{-4} (\eta^{ij} + h^{ij}),
            \end{equation}
            where the conformal factor $\psi$ is given by 
            \begin{equation}
                \psi = \left( \frac{\det \gamma_{ij}}{\det \eta_{ij}} \right)^{1/12}.
            \end{equation}
            By introducing the covariant derivative $\widetilde{D}_i$ associated with the flat metric tensor $\eta_{ij}$ and the corresponding connection
            coefficients $\widetilde{\Gamma}^k_{ij}$, the divergence term in the expansion function (\ref{eq:expansion_function}) is
            \begin{numparts}
            \label{eq:divergent_2}
            \begin{eqnarray}
                \fl
                D_i s^i &= m^{ij} D_i (\lambda m_j) = \lambda m^{ij} D_i m_j
                \\
                \fl
                &= \lambda \psi^{-4} \eta^{ij} \widetilde{D}_i m_j + \lambda \left( \psi^{-4} h^{ij} - s^i s^j \right) \partial_i m_j
                - \lambda (\gamma^{ij} - s^i s^j) \Gamma^k_{ij} \hlmath{m_k} + \lambda \psi^{-4} \eta^{ij} \widetilde{\Gamma}^k_{ij} m_k,
            \end{eqnarray}
            \end{numparts}%
            \hlRthree{where $\Gamma^k_{ij}$ are the connection coefficients associated with the 3-metric $\gamma_{ij}$.}
            One can see that the Laplacian, i.e., the linear elliptic part we are looking for, appears in the first term. In spherical coordinates, it reads
            \begin{equation}
                \eta^{ij} \widetilde{D}_i m_j
                = - \frac{1}{h^2} \left( \Delta_{\theta\phi} h - 2h \right).
            \end{equation}
            The expansion function now becomes
            \begin{numparts}
            \label{eq:expansion_function_full}
            \begin{eqnarray}
                \sqrt{2} \Theta &= D_i s^i - (\gamma^{ij} - s^i s^j) K_{ij}
                \\
                \phantom{\sqrt{2} \Theta}&= - \frac{\lambda \psi^{-4}}{h^2} \left( \Delta_{\theta\phi} h - 2h \right) + \lambda \left( \psi^{-4} h^{ij} - s^i s^j \right) \partial_i m_j
                \nonumber \\
                &\phantom{=} - (\gamma^{ij} - s^i s^j) \left( \lambda \Gamma^k_{ij} \hlmath{m_k} + K_{ij} \right) + \lambda \psi^{-4} \eta^{ij} \widetilde{\Gamma}^k_{ij} m_k.
            \end{eqnarray}
            \end{numparts}%
            Comparing with (\ref{eq:ah_master_propose}), the combination $\Delta_{\theta\phi}h - 2h$ on the right hand side cancels out if we choose $\rho = \sqrt{2} h^2 \psi^{4} / \lambda$, yielding a PDE in suitable form to be solved by relaxation schemes. We make an additional remark that from 
            (\ref{eq:divergent_2}), the linear elliptic operator in the expansion equation is indeed
            \begin{equation}
                h^2 \eta^{ij} \left( \widetilde{D}_i m_j + \widetilde{\Gamma}^k_{ij} m_k \right) = - \left( \partial^2_\theta h + \frac{1}{\sin^2\theta} \partial^2_\phi h \right),
            \end{equation}
            but it turns out that such form is not suitable for relaxation methods to work properly.
            
            Finally, the ansatz we adopt in this work in explicit form is given by
            \begin{equation}
            \label{eq:ah_master_final}
                \Delta_{\theta\phi}h - (2 - \eta) h = S ( h, \partial_i h, \partial^2 h; \gamma_{ij}, \partial_k \gamma_{ij}, K_{ij}; \eta),
            \end{equation}
            with the source term $S$ being
            \begin{numparts}
            \label{eq:ah_master_source}
            \begin{eqnarray}
                S &= \Delta_{\theta\phi} h - (2 - \eta) h + \frac{h^2 \psi^4}{\lambda} (\sqrt{2} \Theta )
                \\
                &= h^2 \psi^4 \bigg[ \left( \psi^{-4} h^{ij} - s^i s^j \right) \partial_i m_j 
                - \left( \gamma^{ij} - s^i s^j \right) \left( \Gamma^k_{ij} m_k + \frac{K_{ij}}{\lambda} \right) \bigg]
                \nonumber \\
                &\phantom{=} + h^2 \eta^{ij} \widetilde{\Gamma}^k_{ij} m_k + \eta h,
            \end{eqnarray}
            \end{numparts}%
            where the extra parameter $\eta$ can speed up convergence in the relaxation method if appropriately chosen \cite{Shibata1997}. \hlRone{Our experience with the benchmark tests (see Section~{\ref{sec:3-tests}}) suggests that the optimal value of $\eta$ can only be found empirically, but in general it should take a value between 0 and 1.}

    \subsection{Implementation of Multigrid Apparent Horizon Finder}
        \subsubsection{Multigrid method}
            Iterative schemes have been a standard tool for solving general elliptic PDEs numerically. For iterative schemes that use a finite difference discretization setup, there is sometimes a trade-off between choosing a grid of higher resolution for more accurate results and a grid of relatively lower resolution for faster convergence. The persistent low-frequency components in the numerical error is one of the factors that reduce the convergent rate, but they can be eliminated rather efficiently at lower resolutions.
            
            A multigrid scheme combines multiple levels of grid with different resolutions to tackle low-frequency components in the error in the coarser grids while retaining good accuracy of the solution in the finest grid. The major components of a multigrid scheme are as follows. 
            \textit{Smoothers} (e.g., standard relaxations) are applied on each but the coarsest level to eliminate error modes at different spectral range, and a \textit{solver} is employed at the coarsest level at which an exact/approximate solution can be found more efficiently. With multiple levels of grid, the inter-grid data transfer operators---\textit{restrictions} and \textit{prolongations}---are there to bridge successive grid levels by transferring the current solutions and/or error corrections between them. Finally, a \textit{cycling scheme} is chosen to specify the exact scheduling, when to jump between different levels of grid, of a multigrid solver. We refer the reader to \cite{briggs2000,Trottenberg2000,Hackbusch1985} for detailed discussion on the multigrid methods.
            
        \subsubsection{Linear multigrid algorithm}
            Owing to the fact that the principle Laplacian in (\ref{eq:ah_master_final}) is linear, we choose to use a linear multigrid algorithm, which will be briefly outlined in the following.
            
            Suppose we solve a linear elliptic equation $\mathcal{L}(u)=f$ on a uniform grid of 
            spacing $k$, and we write
            \begin{equation}
            \label{eq:mglin_poisson}
                \mathcal{L}_k (u_k) = f_k,
            \end{equation}
            where $\mathcal{L}$ is the elliptic operator, $f$ is the source term and $u$ is the exact solution. Let $\widetilde{u}_k$ denote the intermediate approximate solution, and $e_k = u_k - \widetilde{u}_k$ denote the corresponding error. Since $\mathcal{L}$ is linear, (\ref{eq:mglin_poisson}) becomes
            \begin{equation}
            \label{eq:mglin_residual_fine}
                \mathcal{L}_k (e_k) = f_k - \mathcal{L}_k (\widetilde{u}_k) \equiv r_k,
            \end{equation}
            where $r_k$ is called the \textit{residual}. This residual equation is easier to solve on a coarser grid of spacing $2k$ if we consider an approximation version of it, i.e.,
            \begin{equation}
            \label{eq:mglin_residual_coarse}
                \mathcal{L}_{2k} (e_{2k}) = r_{2k},
            \end{equation}
            where the coarser grid residual $r_{2k}$ is found by restricting the finer grid residual $r_k$ using the restriction operator $\mathcal{R}$, i.e., $r_{2k} = \mathcal{R} (r_k)$. The solution to (\ref{eq:mglin_residual_coarse}) can be thought as the correction to the approximate solution $\widetilde{u}_k$ we have for the original problem (\ref{eq:mglin_poisson}). Now denote the approximate solution to (\ref{eq:mglin_residual_coarse}) by $\widetilde{e}_{2k}$, we can interpolate it to the finer grid by the prolongation operator $\mathcal{P}$ such that the approximate solution $\widetilde{u}_k$ is updated by
            \begin{equation}
            \label{eq:mglin_update}
                \widetilde{u}_k^{\rm new} = \widetilde{u}_k + \mathcal{P} (\widetilde{e}_{2k}).
            \end{equation}
            
            The above procedure is an illustration on a 2-grid structure and can be easily generalized for
            an $n$-grid solver. In particular, coarser grids can be recursively constructed when solving for the correction in (\ref{eq:mglin_residual_coarse}).

        \subsubsection{Grid discretization}
            \hlRone{We discretize the apparent horizon surface $h(\theta,\phi)$ using a \textit{vertex-centered grid}. At the finest grid level, there are $N_\theta$ points along the polar direction and $N_\phi$ points along the azimuthal direction.}
            The grid points sit at 
            \begin{numparts}
            \begin{eqnarray}
                \theta_i = i \Delta \theta = i \left( \frac{\pi}{N_\theta - 1} \right), &\qquad i=0,\dots, N_\theta -1;
                \\
                \phi_j = j \Delta \phi = j \left( \frac{2\pi}{N_\phi - 1} \right), &\qquad j=0,\dots, N_\phi-1.
            \end{eqnarray}
            \end{numparts}%
            \hlRone{Note that we require the numbers of grid points in the finest grid level, i.e., $N_\theta$ and $N_\phi$, to be odd number, but this condition can be relaxed for the subsequent coarser grid levels.}
            We further discretize the apparent horizon equation (\ref{eq:ah_master_final}) using a 4th order central finite difference representation for the Laplacian such that it is in a suitable form for conventional relaxation methods.

        \subsubsection{Boundary condition}
            Special treatments are carried out to the boundary points of this vertex-centered spherical grid. 
            \hlRone{In the azimuthal direction, periodic boundary conditions are applied. The points at $\phi_j=2\pi$ are ghost points and their values are set by $h(\theta_i, 2\pi) = h(\theta_i, 0)$. The boundary conditions in the polar direction is a bit more tricky because both} 
            the Laplacian $\Delta_{\theta\phi}$ and the expansion function $\Theta$ exhibit singularity at the poles. The polar singularities are avoided by \emph{not} updating the horizon function $h$ with (\ref{eq:ah_master_final}) there. 
            \hlRone{Instead, we first interpolate one value of $h$ at the (north) pole, say at $(\theta=\theta_0=0,\phi=\phi_k)$ for some $k$, through a cubic polynomial constructed using the points $h(\theta_{-2},\phi_k)$, $h(\theta_{-1},\phi_k)$, $h(\theta_{1},\phi_k)$ and $h(\theta_{2},\phi_k)$, where the negative indices of $\theta$ represent points on the opposite side across the (north) pole. This value of $h(0,\phi_k)$ is then copied to all other points at $\theta=0$, that is, all $h(0,\phi_j)$ are forced to be the same for each $j$. In practice, the value of $k$ does not matter much and we have set $k=0$, i.e., $\phi_k=0$, for convenience. The same procedure is done at the south pole to ensure the smoothness of the surface across both poles.}

        \subsubsection{Smoothers and inter-grid transfer operators}
            In the multigrid algorithm, we need to specify a way to obtain approximate solutions to (\ref{eq:mglin_poisson}) and (\ref{eq:mglin_residual_coarse}). We use the \mbox{Gauss-Seidel} line relaxation method \cite{num_recipe2007} in the $\phi$-direction as a smoother (see Section~\ref{sec:2C-compactFD} for more details) at all grid levels including the coarsest level. The smoothing operator applied before restrictions and after prolongations consists of 1 and 2 sweeps of relaxation, respectively. At the coarsest level, the smoothing operator consists of 100 sweeps of relaxation.
        
            For the inter-grid transfer operators, we choose the bi-linear prolongation operator $\mathcal{P}$ and the full-weight restriction operator $\mathcal{R}$. These operators can be represented in the stencil notations (see, e.g., \cite{Wesseling1992}) by
            \begin{equation}
                \mathcal{P} = \frac{1}{4} 
                \left[ \matrix{
                1 & 2 & 1 \cr
                2 & 4 & 2 \cr
                1 & 2 & 1
                } \right], \qquad 
                \mathcal{R} = \frac{1}{16} 
                \left[ \matrix{
                1 & 2 & 1 \cr
                2 & 4 & 2 \cr
                1 & 2 & 1
                } \right].
            \end{equation}
        
        \subsubsection{Cycling algorithm and the solving procedure}
\begin{figure}
            \begin{center}
            \scalebox{0.5}{
                \LARGE
                \begin{tikzpicture}
                \newcommand\rr{0.2}
                \newcommand\hrr{0.2}
                \newcommand\xstep{0.8}
                \newcommand\ystep{2}
                \begin{scope}
                    \draw (11.75*\xstep,2.75*\ystep) -- (12.25*\xstep,3.25*\ystep);
                    \draw (13*\xstep,3*\ystep) node[right] {Prolongation};
                    \draw (11.75*\xstep,2.25*\ystep) -- (12.25*\xstep,1.75*\ystep);
                    \draw (13*\xstep,2*\ystep) node[right] {Restriction};
                    \filldraw[black, fill=black] (12*\xstep,\ystep) circle (\rr) node{};
                    \draw (13*\xstep,\ystep) node[right] {Smoother};
                    \filldraw[black, fill=white] (12*\xstep-\hrr,-\hrr) rectangle (12*\xstep+\hrr,\hrr);
                    \draw (12*\xstep,0) node {};
                    \draw (13*\xstep,0) node[right] {Exact solver};

                    \draw[loosely dashed] (-2*\xstep,3*\ystep) -- (8*\xstep,3*\ystep);
                    \draw[loosely dashed] (-2*\xstep,2*\ystep) -- (8*\xstep,2*\ystep);
                    \draw[loosely dashed] (-2*\xstep,1*\ystep) -- (8*\xstep,1*\ystep);
                    \draw[loosely dashed] (-2*\xstep,0) -- (8*\xstep,0);

                    \draw (0,3*\ystep) -- (3*\xstep,0);
                    \draw (3*\xstep,0) -- (6*\xstep,3*\ystep);

                    \draw (-2*\xstep,3*\ystep) node[left] {Finest grid};
                    \draw (-2*\xstep,0) node[left] {Coarsest grid};

                    \foreach \i in {0,1,2} 
                        \filldraw[black, fill=black] (\i*\xstep,3*\ystep-\i*\ystep) circle (\rr) node{};
                    \filldraw[black, fill=white] (3*\xstep-\hrr,-\hrr) rectangle (3*\xstep+\hrr,\hrr);
                    \draw (3*\xstep,0) node {};
                    \foreach \i in {4,5,6}
                        \filldraw[black, fill=black] (\i*\xstep,\i*\ystep-3*\ystep) circle (\rr) node{};
                \end{scope}
                \end{tikzpicture}
            }
            \end{center}
            \caption{The structure of one $V$-cycle with 4 grid levels. The operators are applied in chronological order from left to right. In our code, we use the smoother instead of an exact solver at the coarsest grid as well.}
            \label{fig:Vcycle}
            \end{figure}
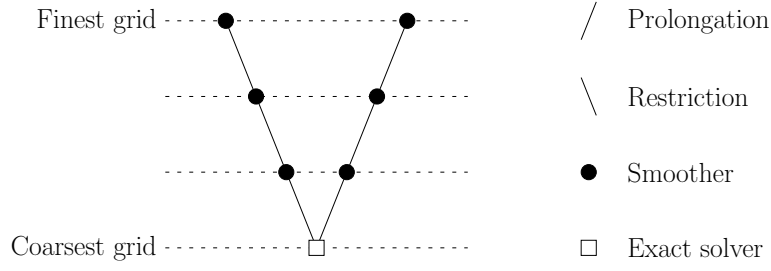

            Among different multigrid cycling algorithms, we choose to use the $V$-cycle (see Figure~\ref{fig:Vcycle}) owing to its simplicity. Successive $V$-cycles are carried out until some prescribed tolerance is satisfied to locate the apparent horizon. 
           
            Assuming that the geometric objects $\gamma_{ij}$ and $K_{ij}$ are given on a hypersurface, an initial trial surface $h^{(0)}$ is specified for the calculation of the source term $S$ according to (\ref{eq:ah_master_source}). Note that we do not need the value of $S$ at the poles since these points are excluded from the relaxation domain. One $V$-cycle is then performed to obtain a new guess surface $h^{(1)}$ while keeping the source term fixed. The new surface $h^{(1)}$ is used to update the source term $S$ before entering the next $V$-cycle. This process is repeated until the maximum change in $h$ between successive trial surfaces, denoted by $\|\delta h\|_\infty$, 
            and/or the maximum value of the expansion on the final guess surface, denoted by $\|\Theta\|_\infty$, are less than some tolerance $\epsilon_h$ and $\epsilon_\Theta$, respectively, within the relaxation domain (i.e., excluding the poles).

        \subsubsection{Handling of numerical spacetimes}
            The ability to process numerical spacetime is important for an apparent horizon finder to be applicable in numerical simulations, where the geometric objects $\gamma_{ij}$ and $K_{ij}$ are given only at discrete locations. To update the source term (\ref{eq:ah_master_source}) between each $V$-cycle with numerical spacetime data, we need the values (in spherical basis) of $\gamma_{ij}$, $K_{ij}$ and also $\partial_k \gamma_{ij}$ on the current trial surface $h(\theta,\phi)$. The current implementation of the interpolator in our finder takes input of $\gamma_{ij}$ and $K_{ij}$ on a uniform Cartesian grid, interpolates $\gamma_{ij}$, $\partial_k \gamma_{ij}$ and $K_{ij}$ in Cartesian basis onto the trial surface using tricubic Hermite spline, and finally transforms the components from Cartesian basis to spherical polar basis. In approximating the first to third order derivatives of $\gamma_{ij}$ and $K_{ij}$ in Cartesian basis which are required for tricubic Hermite interpolation, a standard 4th order finite differencing scheme is used to maintain an overall third order accuracy in the interpolator.  

    \subsection{Line Relaxation Smoothing Operator}
    \label{sec:2C-compactFD}
        As mentioned in the previous section, we have chosen the Gauss-Seidel line relaxation in the azimuthal direction as the smoothing operator in our multigrid code. Due to the inherent asymmetry 
        between the polar and azimuthal directions of the spherical Laplacian $\Delta_{\theta\phi}$, the line relaxation has better convergence, especially near the poles, than a simple point-wise relaxation with or without red-black ordering updates \cite{Barros1991}. Because of the periodic boundary condition in the azimuthal direction, one step of the $\phi$-line relaxation requires solving a cyclic banded diagonal system \cite{num_recipe2007}. In particular, a cyclic penta-diagonal system is needed to be solved for each relaxation step if we use a standard 4th order 5-point finite difference formula. This obviously requires much more work than solving a tri-diagonal system, which is the case with a 2nd order 3-point finite difference scheme. In order to take advantage of the efficiency in solving tri-diagonal system while achieving 4th order accuracy, we aim for a compact finite difference scheme for (\ref{eq:ah_master_final}) (see e.g., \cite{HOC_Zhang2002} and \cite{HOC_Britt2010} respectively for such treatment in Cartesian Poisson equation and polar Helmholtz equation).

        \hlRone{The compact finite difference scheme, or the Mehrstellenverfahren discretization (e.g., {\cite{Trottenberg2000}}) is not a new concept and can be traced back to as early as the 1960's {\cite{Collatz1960}}, which was then referred to as the Hermitian method. It%
        }
        aims at increasing the accuracy of a finite difference scheme without involving more grid points that are \textit{too} far away from the current pivoting point. As an example, for the Poisson equation $(\partial_x^2 + \partial_y^2)u = f$ on a square Cartesian grid of step size $k$, the traditional 4th order \hlRone{(non-compact)} finite difference scheme involves the second next neighboring grids, as represented by the stencil notation
        \begin{equation*}
            \frac{1}{12k^2}
            \left[ \matrix{
            && -1 && \cr
            && 16 && \cr
            -1&16&-60&16&-1 \cr
            && 16 && \cr
            && -1 && 
            } \right] u_{ij}
            =
            f_{ij},
        \end{equation*}
        while its 4th order compact finite difference representation is \cite{Collatz1960}
        \begin{equation*}
            \frac{1}{6k^2} 
            \left[ \matrix{
            1 & 4 & 1 \cr
            4 & -20 & 4 \cr
            1 & 4 & 1
            } \right] u_{ij}
            =
            \frac{1}{12} 
            \left[ \matrix{
            0 & 1 & 0 \cr
            1 & 8 & 1 \cr
            0 & 1 & 0
            } \right] f_{ij}.
        \end{equation*}
        Notice the smaller stencil size (width) in the compact finite difference scheme. This compact representation can be viewed as the incorporation of a correction to the 2nd order finite difference formula to eliminate the leading error in its Taylor's expansion (see, e.g., \cite{HOC_Zhang2002,HOC_Lai2007,HOC_Ge2010,HOC_Baruch2009,HOC_Britt2010,HOC_Deriaz2020} for a variety of PDEs that such method has been applied to). 
        The compact form for the spherical Poisson equation, to which the apparent horizon equation (\ref{eq:ah_master_final}) belongs, can be derived in a similar manner. 
        We refer the reader to \cite{HOC_Deriaz2020} for a summary of compact scheme for Poisson equation on Cartesian grids.
        
        Since our ultimate goal is to create a stencil that is suitable for reducing the $\phi$-line relaxation to solving a tri-diagonal system, we only need to ``squeeze'' the usual 5-point-wide stencil in the $\phi$ direction, which we refer to as \textit{semi-compact}. 
        We derive such \textit{semi-compact} stencil for a general 2-dimensional spherical Poisson equation in the following\footnote{%
            For completeness, we also derive the \textit{fully}-compact representation in \ref{appendix:compactFD} for a general 2-dimensional Poisson equation on the sphere. We do not use the \textit{fully}-compact form because it requires the value of the source term $f$ at the poles, but the apparent horizon equation (\ref{eq:ah_master_final}) contains the expansion function $\Theta$ which is undefined at the poles in spherical coordinates (see (\ref{eq:expansion_function_full})).}. 
        Suppose we solve
        \begin{equation}
        \label{eq:linerelax_master}
            \partial^2_\theta u + \cot\theta \,\partial_\theta u + \frac{1}{\sin^2\theta} \partial^2_\phi u = f(\theta, \phi).
        \end{equation}
        First observe that using a 2nd order finite difference method for the third term, we have
        \begin{equation}
        \label{eq:linrelax_eq1}
            \frac{1}{\sin^2\theta} \pdv[2]{u}{\phi} 
            = \left( \frac{1}{\sin^2\theta} \pdv[2]{u}{\phi} \right)_{\rm FD2} - \frac{k^2}{12\sin^2\theta} \pdv[4]{u}{\phi} + \mathcal{O}(k^4),
        \end{equation}
        where the subscript FD$n$ indicates that the corresponding term is approximated by an $n$-th order finite difference scheme, and $k$ is the step size in the $\phi$ direction. By differentiating (\ref{eq:linerelax_master}) twice w.r.t $\phi$ and substituting the result to (\ref{eq:linrelax_eq1}), we obtain
        \begin{equation}
        \label{eq:compactFD_phi_deriv}
            \fl
            \frac{1}{\sin^2\theta} \pdv[2]{u}{\phi} 
            = \left( \frac{1}{\sin^2\theta} \pdv[2]{u}{\phi} \right)_{\rm FD2}
            -\frac{k^2}{12} \left[ \pdv[2]{f}{\phi} - \frac{\partial^4u}{\partial\theta^2\partial\phi^2} - \cot\theta \frac{\partial^3u}{\partial\theta\partial\phi^2} \right] + \mathcal{O}(k^4).
        \end{equation}
        With a factor of $k^2$ in front, the terms inside the square bracket only need to be evaluated at 2nd order accuracy to maintain an overall 4th order accuracy. Finally, assuming that the step size in $\theta$ direction is also $k$ for simplicity, the \textit{semi-compact} 4th order finite difference representation of (\ref{eq:linerelax_master}) can be given by
        \begin{eqnarray}
            \left( \pdv[2]{u}{\theta} + \cot\theta \pdv{u}{\theta} \right)_{\rm FD4} + \left( \frac{1}{\sin^2\theta} \pdv[2]{u}{\phi} \right)_{\rm FD2} &
            \nonumber \\
            -\frac{k^2}{12} \left[ \pdv[2]{f}{\phi} - \frac{\partial^4u}{\partial\theta^2\partial\phi^2} - \cot\theta \frac{\partial^3u}{\partial\theta\partial\phi^2} \right]_{\rm FD2} + \mathcal{O}(k^4)
            &= f.
        \end{eqnarray}
        The adaption of this form to our apparent horizon equation (\ref{eq:ah_master_final}) is straight-forward, despite the fact that the source term now also depends on the solution. Notice that the stencil is only 3-point-wide in the $\phi$ direction such that a $\phi$-line relaxation can be done by a more efficient tri-diagonal algorithm. 
        
        \begin{table}
        \caption{\label{tab:line_relax_efficiency}
        Comparison of the efficiency between using point-wise relaxation and line-relaxation as the smoothing operator to locate the apparent horizon for the analytic Kerr-Schild spacetime of a black hole with mass $M=1$ and spin parameter $a=0.8$. The columns represent the resolution, number of grid levels $n$, and the number of $V$-cycle and user CPU time (in seconds) required to reach the solution for both relaxation methods. The tolerance parameters are set to $\epsilon_h = \epsilon_\Theta = 10^{-8}$. The results in the rows marked with (without) an asterisk are obtained using $\eta=0.8$ (0).
        \hlRone{For the pointwise relaxation, the number of relaxation sweeps before restrictions and after prolongation is changed to 10 and 20, and that at the coarsest level to 200. Both the results of pointwise relaxation using the non-compact (NC) and the semi-compact (SC) finite difference scheme are shown for comparison. }}
        \begin{indented}
        \item[]\begin{tabular}{@{}lccrcrcr}
        \br
            &&
            \centre{2}{point relax.~(NC)}
            &\centre{2}{point relax.~(SC)}
            &\centre{2}{line relaxation}\\
        \ns\ns
        &&\crule{2}&\crule{2}&\crule{2}\\
            $N_\theta{\times}N_\phi$ &$n$-grid 
            &$V$-cycle &Time(s) 
            &$V$-cycle &Time(s) 
            &$V$-cycle &Time(s)\\
        \mr
        \multirow{2}*{{\medmuskip=0mu $65\times 65$}} 
            & 3 & 36 & 0.251 &35 &0.330 & 34 & 0.204\\
            &\phantom{$^*$}3$^*$ & 22 & 0.161 &20 &0.194 & 19 & 0.122\\
        \mr
        \multirow{3.5}*{{\medmuskip=0mu $129\times 129$}}
            &3 &80 &2.170 &66 &2.671 &45 &1.035\\
            \cmidrule{2-8}
            &4 &64 &1.652 &53 &1.794 &34 &0.735\\
            &\phantom{$^*$}4$^*$ &67 &1.743 &54 &1.822 &20 &0.446\\
        \mr
        \multirow{5.5}*{{\medmuskip=0mu $257\times 257$}}
            &3 &318 &33.502 &254 &37.065 &128 &11.537 \\
            \cmidrule{2-8}
            &4 &272 &27.828 &217 &28.664 &46 &3.903 \\
            &\phantom{$^*$}4$^*$ &278 &28.332 &219 &29.177 &35 &3.009 \\
            \cmidrule{2-8}
            &5 &260 &25.884 &205 &26.693 &34 &2.857 \\
            &\phantom{$^*$}5$^*$ &257 &26.624 &204 &26.529 &21 &1.814 \\
        \br
        \end{tabular}
        \end{indented}
        \end{table}
        
        We make some final remarks to conclude this section. First, a simple point-wise Gauss-Seidel relaxation 
        \hlRone{(using either 5-point-wide non-compact or semi-compact finite difference scheme)} can certainly serve as the smoothing operator, but the efficiency of the solver will be much degraded (See Table \ref{tab:line_relax_efficiency} and the discussion in Section \ref{sec:3.1-KS}).
        \hlRone{It is worth noting that although using semi-compact scheme gives a slightly better convergence (in terms of the number of $V$-cycles) than using the usual non-compact scheme, the computational costs of the two are roughly the same due to the more arithmetic operations needed in the semi-compact scheme.}
        Second, the asymmetry in the spherical Laplacian may change direction from the pole to the equator depending on the grid sizes in the $\theta$- and $\phi$-directions \cite{Barros1991}. In the case when the number of points in the $\theta$-direction is much larger than that in the $\phi$-direction, the $\phi$-line relaxation may not provide good convergence, especially near the equator, or it may even fail. A remedy is to use a combined relaxation \cite{Barros1991}, sometimes referred to as the segment relaxation scheme, which applies the $\theta$- and $\phi$-line relaxations to different domains according to the direction of asymmetry in the Laplacian. Although this could be relevant for further increasing the efficiency of our finder when applied to spacetimes with axisymmetry, we will continue to use the $\phi$-line relaxation here.

\section{Results}
\label{sec:3-tests}
    We have tested our multigrid apparent horizon finder with several benchmark spacetimes and the results are reported in this section. All tests are run on a \SI{3.6}{GHz} processor and the timings are reported in user CPU time.
    
    The geometric objects $\gamma_{ij}$ and $K_{ij}$ are given analytically in spherical coordinates for the calculation of the source term $S$ (\ref{eq:ah_master_source}), except for numerical data where they are interpolated onto the surface. Unless otherwise specified, the initial guess surface is a sphere of coordinate radius $1.5M$, where $M$ is the mass of the black hole. The tolerance to stop the iteration process is set to $\epsilon_h = \epsilon_\Theta = 10^{-8}$, and the convergence parameter is set to $\eta=0.8$. After locating the apparent horizon, its proper area is calculated numerically by \cite{Lin2007}
    \begin{eqnarray}
        A = &\int_0^{2\pi} \rmd \phi \int_0^{\pi} \rmd \theta \,\bigg[ \left( \gamma_{rr} h_{,\theta}^2 + 2\gamma_{r\theta} h_{,\theta} + \gamma_{\theta\theta} \right) \left( \gamma_{rr} h_{,\phi}^2 + 2\gamma_{r\phi} h_{,\phi} + \gamma_{\phi\phi} \right) 
        \nonumber \\
        &- \left( \gamma_{rr} h_{,\theta} h_{,\phi} + \gamma_{r\phi} h_{,\theta} + \gamma_{r\theta} h_{,\phi} + \gamma_{\theta\phi} \right)^2 \bigg]^{1/2}.
    \end{eqnarray}

    \subsection{Kerr-Schild spacetime}
    \label{sec:3.1-KS}
        For describing a Kerr black hole of mass $M=1$ and spin parameter $a$ ($0\leq a<1$), one can use the Kerr-Schild coordinates \cite{MTW1973gravitation}, of which the spatial metric $\gamma_{ij}$ and the extrinsic curvature $K_{ij}$ are given in Cartesian coordinates by
        \begin{numparts}
        \begin{eqnarray}
            \gamma_{ij} &= \eta_{ij} + 2H \ell_i \ell_j;
            \\
            K_{ij} &= 2\alpha H \ell^k \partial_k (H \ell_i \ell_j) + \alpha [ \partial_i (H \ell_j) + \partial_j (H\ell_i) ],
        \end{eqnarray}
        \end{numparts}%
        where $\eta_{ab} = {\rm diag}(-1,1,1,1)$ is the flat metric and $\alpha \equiv \sqrt{1+2H}$ is the lapse function. 
        The function $H$ and the auxiliary components $\ell^i$ are defined by
            \begin{equation}
                H \equiv \frac{MR^3}{R^4+a^2 z^2};
                \qquad
                \ell_i = \ell^i \equiv \left( \frac{Rx+ay}{R^2+a^2}, \frac{Ry-ax}{R^2+a^2}, \frac{z}{R} \right),
            \end{equation} 
            with the parameter $R$ being the Boyer-Lindquist radial coordinate which satisfies
            \begin{equation}
                \frac{x^2+y^2}{R^2+a^2} + \frac{z^2}{R^2} = 1.
            \end{equation}
        The analytic forms of $\gamma_{ij}$, $\partial_k \gamma_{ij}$ and $K_{ij}$, after a coordinate transformation, serve as the input for the calculation of the source term $S$ in (\ref{eq:ah_master_final}). There are two horizons at $R=r_\pm\equiv M(1\pm \sqrt{1-a^2})$ and the apparent horizon is given by the outer one, which is, in spherical coordinates $(r, \theta, \phi)$,
        \begin{equation}
            r^2 = \frac{r_+^2 ( r_+^2 + a^2)}{r_+^2 + a^2 \cos^2\theta},
        \end{equation}
        with the corresponding proper area given by 
        \begin{equation}
            A_{\rm Kerr} = 4\pi (r_+^2 + a^2).
        \end{equation}

        Before considering the performance of our apparent horizon finder, we first use the analytic 
        Kerr-Schild spacetime as a test case to show the advantage of using a line relaxation smoother over a point-wise one in a multigrid code as noted in Section~\ref{sec:2C-compactFD}. 
        As shown in Table~\ref{tab:line_relax_efficiency}, the number of $V$-cycles required to reach the solution is less when a line relaxation is used. The difference becomes more significant with higher resolution and more grid levels. Combined with the enhancement given by tuning the convergence parameter $\eta$, the use of line relaxation over point relaxation leads to an order-of-magnitude  
        overall boost in terms of speed in certain test cases.
        
        We use several resolutions to locate the apparent horizon in Kerr-Schild spacetime with spin parameter $a=0.6$ and 0.9. The results are reported in Table \ref{tab:KS_a0.6_a0.9}. With the same number of grid levels, increasing the resolution slightly increases the number of iteration needed to reach convergence. This, however, can be compensated by introducing more grid levels. Overall, we find that the number of iterations ($V$-cycles) to reach convergence is more or less the same for a fixed $a$, independent of the resolutions. The run time reported here is only determined by the searching algorithm, excluding the interpolation routines required in numerical spacetimes (see Section~\ref{sec:num_spacetimes}). We find that the run time is proportional to the total number of grid points on the finest grid level.

        \begin{table}
        \caption{\label{tab:KS_a0.6_a0.9} 
            Finding the apparent horizon in analytic Kerr-Schild spacetime of spin parameter 
            $a=0.6$ and 0.9. The columns represent the resolution \tightres{N_\theta}{N_\phi}
            on the finest grid level, the number of grid levels $n$; the relative error in the proper area $\delta A/A_{\rm Kerr}$; the number of $V$-cycle and the user CPU time (in seconds) required to locate the apparent horizon. 
            In the first column, the angle (e.g., \SI{5}{\degree}) inside the parenthesis 
            in each row is approximately the angular separation between two neighboring grid points in the $\theta$- or $\phi$-direction.
        }
        \begin{indented}
        \item[]\begin{tabular}{@{}ccccrccr}
        \br
            & & \centre{3}{$a=0.6$} & \centre{3}{$a=0.9$}\\
            \ns\ns & & \crule{3} & \crule{3} \\
            \tightres{N_\theta}{N_\phi} &$n$-grid & $|\delta A/A_{\rm Kerr}|$ &\#$V$ &Time (s)
            & $|\delta A/A_{\rm Kerr}|$ &\#$V$ &Time (s) \\
        \mr
            \enspace\tightres{37}{73}\enspace\,(\SI{5}{\degree}) & 4 & \tightnum{2.8e-6} & 16 & 0.063 & \tightnum{1.2e-6} & 25 & 0.096 \\
        \mr
            \enspace\tightres{61}{121}\,(\SI{3}{\degree}) & 4 & \tightnum{3.6e-7} & 16 & 0.162 & \tightnum{1.7e-7} & 26 & 0.249 \\
            \enspace\tightres{61}{121}\,(\SI{3}{\degree}) & 5 & \tightnum{3.6e-7} & 16 & 0.160 & \tightnum{1.7e-7} & 25 & 0.243 \\
        \mr
            \enspace\tightres{91}{181}\,(\SI{2}{\degree}) & 4 & \tightnum{7.2e-8} & 17 & 0.377 & \tightnum{3.9e-8} & 26 & 0.562 \\
            \enspace\tightres{91}{181}\,(\SI{2}{\degree}) & 5 & \tightnum{7.2e-8} & 16 & 0.350 & \tightnum{4.2e-8} & 25 & 0.525 \\
            \enspace\tightres{91}{181}\,(\SI{2}{\degree}) & 6 & \tightnum{7.2e-8} & 16 & 0.349 & \tightnum{4.0e-8} & 24 & 0.510 \\
        \br
        \end{tabular}
        \end{indented}
        \end{table}

    \subsection{Brill-Lindquist spacetime}
        \newcommand{\rbar}{\bar{r}}
        The second test for our finder is to locate the apparent horizon in Brill-Lindquist multiple black hole data \cite{Brill_Lindquist1963}. 
        \hlRone{This is a classic analytic solution for the constraint equations in 3+1 numerical relativity that represents a multiple black hole spacetime. The black holes are assumed to be momentarily at rest. 
        The derivation of this solution can be found in standard numerical relativity textbooks 
        (e.g., {\cite{Baumgarte_Shapiro_2010})}. }
        Under time symmetry and the assumption that the spatial slice is conformally flat, the 3-metric and the extrinsic curvature for $N$-black-hole spacetime in isotropic coordinates $(\rbar, \theta, \phi)$ are, respectively,
        \begin{numparts}
        \begin{eqnarray}
            \gamma_{ij} &= \left( 1+ \sum_{\alpha=1}^N \frac{M_\alpha}{2 \vert \vb*{\rbar}-\vb*{\rbar}_\alpha \vert} \right)^4 {\rm diag}(1, \rbar^2, \rbar^2 \sin^2\theta ), 
            \\
            K_{ij}&=0,
        \end{eqnarray}
        \end{numparts}%
        where $M_\alpha$ and $\vb*{\rbar}_\alpha$ are the mass at infinite separation and the coordinate position of the $\alpha$-th black hole, respectively. 
        \hlRone{The isotropic coordinates here are the analog of those in the isotropic Schwarzschild metric, which can be obtained by substituting the radial coordinate in the Schwarzschild metric by $r=\rbar [1+M/(2\rbar)]^2$.}
        In the following, we consider the case of $N=2$ and $N=3$. We use very high resolution of {\medmuskip=0mu $N_\theta\times N_\phi=385\times769$} (angular spacing $<\SI{0.5}{\degree}$) with 6 levels of grid in these tests without assuming any symmetry of the apparent horizon. We relax the convergence criteria in this section to $\epsilon_h=\num{e-8}$ and $\epsilon_\Theta=\num{e-6}$.

        \subsubsection{Brill-Lindquist 2-black-hole spacetime}
            Let us first consider the Brill-Lindquist data for two black holes. 
            When the two black holes are far away from each other, each black hole possesses its own apparent horizon. With decreasing separation distance $d$, there exists a critical separation $d=d_{\rm crit}$ at which the two black holes start to have a common apparent horizon. One of the benchmark tests is to determine the value of $d_{\rm crit}$. The center of the two black holes are on the $z$-axis at $(0,0,\pm d/2)$ in the following.
            
            We first consider two equal-mass black holes with $M_1=M_2=1$. Our finder reports the critical separation to be $d_{\rm crit}=1.5323948$, which agrees well with the generally accepted values in the literature (e.g., 1.5323949 \cite{Thornburg2003}, 1.532 \cite{Lin2007}). The critical area is found to be $A_{\rm crit}=196.40795$ (agrees with \cite{Thornburg2003} to 5 decimal places). At the critical separation \hlRtwo{we found}, the maximum expansion on the \hlRtwo{numerically-computed} apparent horizon is {\medmuskip=0mu $\|\Theta\|_\infty = \num{2.3e-8}$}. We also report the results for finding the apparent horizon near the critical separation at $d=1.532\approx d_{\rm crit}$ for lower resolutions in Table \ref{tab:BL2BH}.
            \begin{table}
            \caption{\label{tab:BL2BH}
            Finding the apparent horizon for analytic Brill-Lindquist equal-mass 2-black-hole spacetime at $d=1.532\approx d_{\rm crit}.$ While the spacetime is axisymmetric, we do not impose it in the tests.  }
            \begin{indented}
            \item[]\begin{tabular}{@{}cccr}
            \br
                {\medmuskip=0mu $N_\theta\times N_\phi$} &$A$ &$\|\Theta\|_\infty$ &Time (s) \\
            \mr
                {\medmuskip=0mu $\enspace41\times 73\enspace\,(\SI{5}{\degree})$} & 196.4138 & \tightnum{1.57e-8} & \enspace1.22\\
                {\medmuskip=0mu $\enspace61\times 121\,(\SI{3}{\degree})$} & 196.4158 & \tightnum{1.50e-8} & \enspace1.93\\
                %
                %
                {\medmuskip=0mu $\enspace91\times 181\,(\SI{2}{\degree})$} & 196.4162 & \tightnum{1.52e-8} & \enspace4.16\\
                %
                %
                {\medmuskip=0mu $181\times 361\,(\SI{1}{\degree})$} & 196.4163 & \tightnum{1.57e-8} & 16.20\\
                %
            \br
            \end{tabular}
            \end{indented}
            \end{table}

            We then consider unequal-mass black holes with $M_1=0.2$ and $M_2=0.8$. The initial guess surface for this case is changed to a sphere of coordinate radius $\rbar=1.5$. 
            Our finder reports the critical separation to be $d_{\rm crit}=0.6987161$ with the proper area being $A=49.688602$. The maximum expansion on the 
            \hlRtwo{numerically-computed apparent horizon} is {\medmuskip=0mu $\|\Theta\|_\infty = \num{2.1e-7}$}. The value of the critical separation agrees with that reported recently in \cite{Pook-Kolb2019} by using an axisymmetric apparent horizon finder.

        \subsubsection{Brill-Lindquist 3-black-hole spacetime}
\begin{figure}
                \centering
                \includegraphics[width=.42\linewidth]{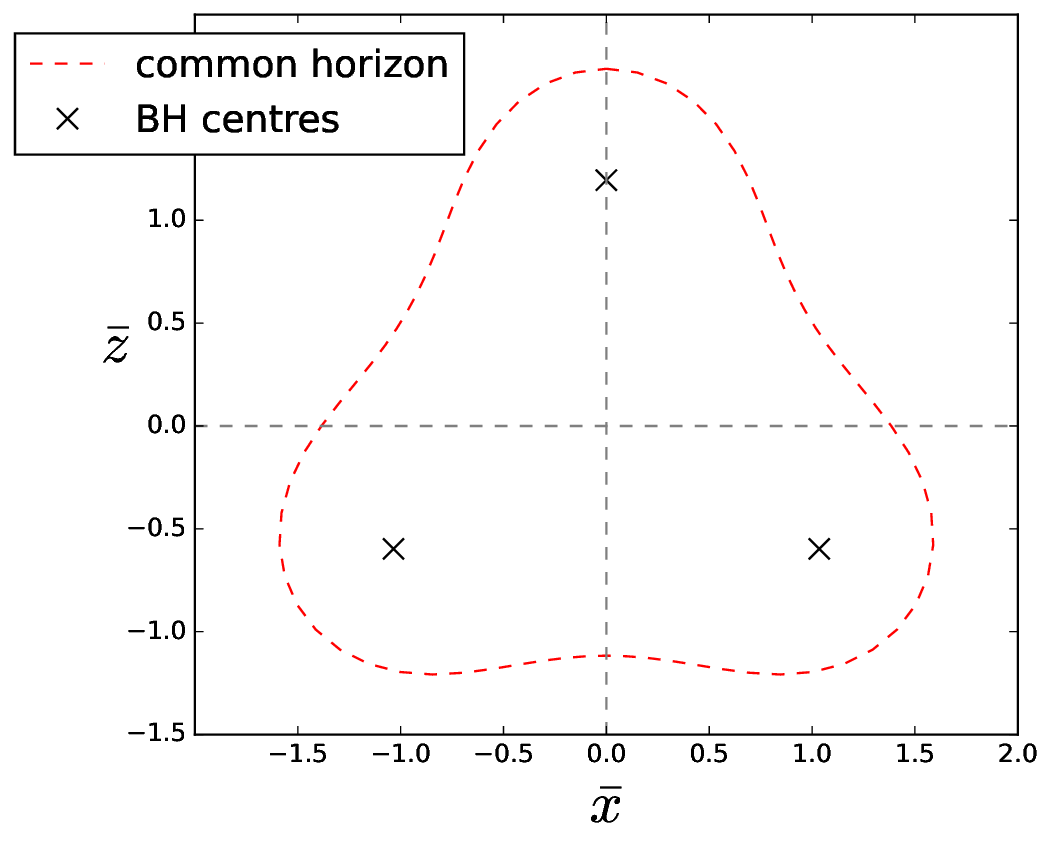}
                \scalebox{.72}{
                  \gdef\gplbacktext{}%
                  \gdef\gplfronttext{}%
                \begingroup
                  \ifGPblacktext
                    \def\colorrgb#1{}%
                    \def\colorgray#1{}%
                  \else
                    \ifGPcolor
                      \def\colorrgb#1{\color[rgb]{#1}}%
                      \def\colorgray#1{\color[gray]{#1}}%
                      \expandafter\def\csname LTw\endcsname{\color{white}}%
                      \expandafter\def\csname LTb\endcsname{\color{black}}%
                      \expandafter\def\csname LTa\endcsname{\color{black}}%
                      \expandafter\def\csname LT0\endcsname{\color[rgb]{1,0,0}}%
                      \expandafter\def\csname LT1\endcsname{\color[rgb]{0,1,0}}%
                      \expandafter\def\csname LT2\endcsname{\color[rgb]{0,0,1}}%
                      \expandafter\def\csname LT3\endcsname{\color[rgb]{1,0,1}}%
                      \expandafter\def\csname LT4\endcsname{\color[rgb]{0,1,1}}%
                      \expandafter\def\csname LT5\endcsname{\color[rgb]{1,1,0}}%
                      \expandafter\def\csname LT6\endcsname{\color[rgb]{0,0,0}}%
                      \expandafter\def\csname LT7\endcsname{\color[rgb]{1,0.3,0}}%
                      \expandafter\def\csname LT8\endcsname{\color[rgb]{0.5,0.5,0.5}}%
                    \else
                      \def\colorrgb#1{\color{black}}%
                      \def\colorgray#1{\color[gray]{#1}}%
                      \expandafter\def\csname LTw\endcsname{\color{white}}%
                      \expandafter\def\csname LTb\endcsname{\color{black}}%
                      \expandafter\def\csname LTa\endcsname{\color{black}}%
                      \expandafter\def\csname LT0\endcsname{\color{black}}%
                      \expandafter\def\csname LT1\endcsname{\color{black}}%
                      \expandafter\def\csname LT2\endcsname{\color{black}}%
                      \expandafter\def\csname LT3\endcsname{\color{black}}%
                      \expandafter\def\csname LT4\endcsname{\color{black}}%
                      \expandafter\def\csname LT5\endcsname{\color{black}}%
                      \expandafter\def\csname LT6\endcsname{\color{black}}%
                      \expandafter\def\csname LT7\endcsname{\color{black}}%
                      \expandafter\def\csname LT8\endcsname{\color{black}}%
                    \fi
                  \fi
                    \setlength{\unitlength}{0.0500bp}%
                    \ifx\gptboxheight\undefined%
                      \newlength{\gptboxheight}%
                      \newlength{\gptboxwidth}%
                      \newsavebox{\gptboxtext}%
                    \fi%
                    \setlength{\fboxrule}{0.5pt}%
                    \setlength{\fboxsep}{1pt}%
                \begin{picture}(7200.00,4540.00)%
                    \gplgaddtomacro\gplbacktext{%
                      \csname LTb\endcsname
                      \put(3275,197){\makebox(0,0){\strut{}$-1$}}%
                      \csname LTb\endcsname
                      \put(4182,534){\makebox(0,0){\strut{}$0$}}%
                      \csname LTb\endcsname
                      \put(5090,871){\makebox(0,0){\strut{}$1$}}%
                      \csname LTb\endcsname
                      \put(2148,255){\makebox(0,0)[r]{\strut{}$-0.5$}}%
                      \csname LTb\endcsname
                      \put(1886,546){\makebox(0,0)[r]{\strut{}$0$}}%
                      \csname LTb\endcsname
                      \put(1624,838){\makebox(0,0)[r]{\strut{}$0.5$}}%
                      \put(1407,1638){\makebox(0,0)[r]{\strut{}$-1$}}%
                      \put(1407,2441){\makebox(0,0)[r]{\strut{}$0$}}%
                      \put(1407,3243){\makebox(0,0)[r]{\strut{}$1$}}%
                      \put(1407,4046){\makebox(0,0)[r]{\strut{}$2$}}%
                    }%
                    \gplgaddtomacro\gplfronttext{%
                      \csname LTb\endcsname
                      \put(4402,332){\makebox(0,0){\strut{}$\bar{x}$}}%
                      \put(1239,349){\makebox(0,0){\strut{}$\bar{y}$}}%
                      \put(1005,2640){\makebox(0,0){\strut{}$\bar{z}$}}%
                    }%
                    \gplbacktext
                    \put(0,0){\includegraphics{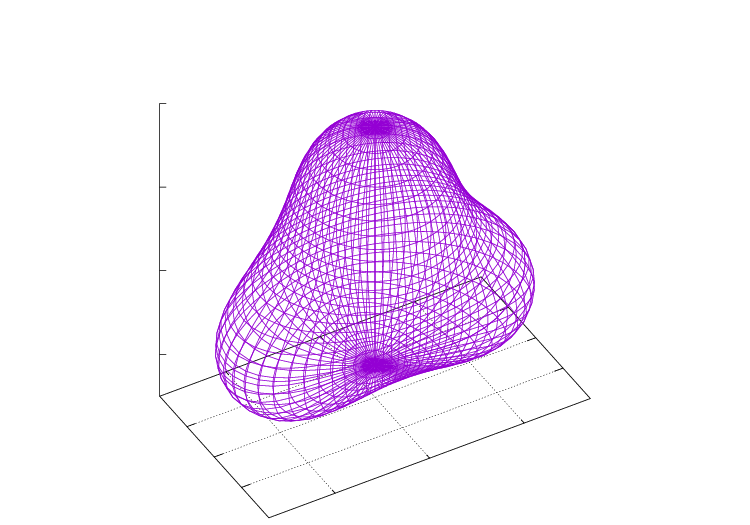}}%
                    \gplfronttext
                  \end{picture}%
                \endgroup
                }
                \caption{The apparent horizon for Brill-Lindquist equal-mass 3-blackhole spacetime at $R=1.195\approx R_{\rm crit}$. The left panel shows the cross-section (red dashed line) of 
                the common horizon on the $\bar{x}$-$\bar{z}$ plane; the right panel is the side view of the 
                full surface of the common horizon.}
                \label{fig:BL3BH}
            \end{figure}
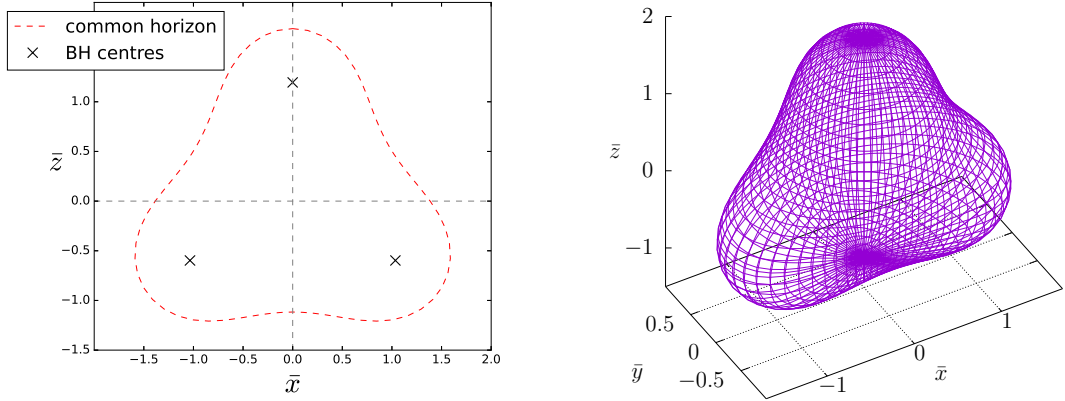

            We now consider a system of three black holes which are placed to form an equilateral triangle on the $\bar{x}$-$\bar{z}$ plane, each at a coordinate distance $R$ from the origin, and one of them sits on the $+\bar{z}$-axis. Each black holes has unity mass. We find that the critical parameter for the three black holes to have a common apparent horizon is $R_{\rm crit} = 1.1954995$, which agrees with the result in \cite{Thornburg2003} to seven decimal places. 
            The critical area and the maximum expansion on the 
            \hlRtwo{numerically-computed apparent horizon} are $A=444.75623$ and {\medmuskip=0mu $\|\Theta\|_\infty = \num{2.3e-8}$}, respectively. 
            The left panel of Figure~\ref{fig:BL3BH} shows the cross section (red dashed line) of the common horizon on the $\bar{x}$-$\bar{z}$ plane for the case $R=1.195 \approx R_{\rm crit}$. 
            The right panel shows the side view of the full surface of the common horizon. 
            
    \subsection{Numerical spacetimes}
    \label{sec:num_spacetimes}
        We prepare numerical spacetimes to further test our finder using the open source code \textsc{Einstein Toolkit} \cite{ET_paper}. The simulation variables $\gamma_{ij}$ and $K_{ij}$ are post-processed to produce data on a uniform rectangular Cartesian grid before importing to our finder. In the case of dynamical binary black hole simulation where adaptive mesh refinement is turned on, we use the python package $\textsc{Kuibit}$ \cite{kuibit} for the post-processing. We also activate the inherent apparent horizon finder in the code, \ahfd\ \cite{Thornburg1996,Thornburg2003}, for comparison. 

        \subsubsection{Numerical Kerr-Schild spacetime}
            We first test our finder with an off-centered Kerr-Schild data. The numerical spacetime data spans the Cartesian grid space of $x,y,z\in [-2.5,2.5]$ with a grid spacing of $\Delta x = \Delta y = \Delta z=0.05$. We use a sphere of coordinate radius $1.5$ centered at the coordinate origin as the initial guess surface. We use the ILUCG matrix routines in all cases for \ahfd. 

            \begin{table}[t]
            \caption{\label{tab:numKS_a0.6}
            Finding the apparent horizon in numerical off-centered Kerr-Schild data with $a=0.6$ (top) and $a=0.8$ (bottom). The first column represents the resolution to describe the horizon surface. The following four columns represent the number of grid levels $n$,
            the relative error in the proper area $\delta A/A_{\rm Kerr}$, 
            the number of $V$-cycle and the time (in seconds) required to locate the apparent horizon using our code. The last two columns represent the number of Newton steps and time (in seconds) required for \ahfd. See text for the numerical setup of the spacetime. 
            }
            \begin{indented}
            \item[]\begin{tabular}{@{}lcccrcr}
            \br
                \quad$a=0.6$ & \centre{4}{\textsc{MGAHF}} & \centre{2}{\ahfd} \\
                \ns\ns &\crule{4}&\crule{2}\\
                \enspace Resolution & $n$-grid & $|\delta A/A_{\rm Kerr}|$ & \#$V$ & Time (s) & \#Newton & Time (s)\\
            \mr
                \enspace \tightres{37}{73}\enspace\,(\SI{5}{\degree}) & 4 & \tightnum{2e-6} & 18 & 2.635 & 8 & 1.604 \\
                \enspace \tightres{61}{121}\,(\SI{3}{\degree}) & 5 & \tightnum{2e-7} & 17 & 6.294 & 8 & 4.180  \\
                \enspace \tightres{91}{181}\,(\SI{2}{\degree}) & 6 & \tightnum{1e-8} & 17 & 12.387 & 8 & 11.393 \\
                \tightres{181}{361}\,(\SI{1}{\degree}) & 7 & \tightnum{1e-8} & 18 & 33.802 & 8 & 88.632 \\
            \br
            \end{tabular}
            \end{indented}
            \begin{indented}
            \item[]\begin{tabular}{@{}lcccrcr}
            \br
                \quad$a=0.8$ & \centre{4}{\textsc{MGAHF}} & \centre{2}{\ahfd} \\
                \ns\ns &\crule{4}&\crule{2}\\
                \enspace Resolution & $n$-grid & $|\delta A/A_{\rm Kerr}|$ & \#$V$ & Time (s) & \#Newton & Time (s)\\
            \mr
                \enspace \tightres{37}{73}\enspace\,(\SI{5}{\degree}) & 4 & \tightnum{2e-6} & 23 & 3.251 & 8 & 1.628 \\
                \enspace \tightres{61}{121}\,(\SI{3}{\degree}) & 5 & \tightnum{3e-7} & 22 & 7.864 & 8 & 4.170 \\
                \enspace \tightres{91}{181}\,(\SI{2}{\degree}) & 6 & \tightnum{4e-8} & 21 & 14.461 & 8 & 11.009 \\
                \tightres{181}{361}\,(\SI{1}{\degree}) & 7 & \tightnum{1e-7} & 21 & 37.178 & 8 & 87.173 \\
            \br
            \end{tabular}
            \end{indented}
            \end{table}

            We consider a black hole of unity mass and spin $a$ that is located at $(0.2,0.2,0.2)$ in Cartesian coordinates. We tabulate the results for $a=0.6$ and $a=0.8$ in Table \ref{tab:numKS_a0.6}. The time to locate the solution surface mainly depends on two factors---the overall convergence of the algorithm and the computational cost within each iteration. The convergence order of the algorithm is important in the sense that as many as 30 interpolations for the geometric variables are required at every grid point at each (outer) iteration. Since we adopt a pure multigrid algorithm, which has a slower convergence than the quadratic convergence in Newton's method, one can see that our finder needs more iterations (number of $V$-cycles) for convergence, so more time is spent on the interpolation routines when compared to \ahfd. Notice that the number of iterations is insensitive to the resolution used for fixed $a$. Had we used the pointwise Gauss-Seidel relaxation as the smoothing operator, the number of iterations would grow linearly with the number of grid points, and the speed of the code would become very slow due to the large amount of interpolations required.

\begin{figure}
                \centering
                \includegraphics[width=.8\textwidth]{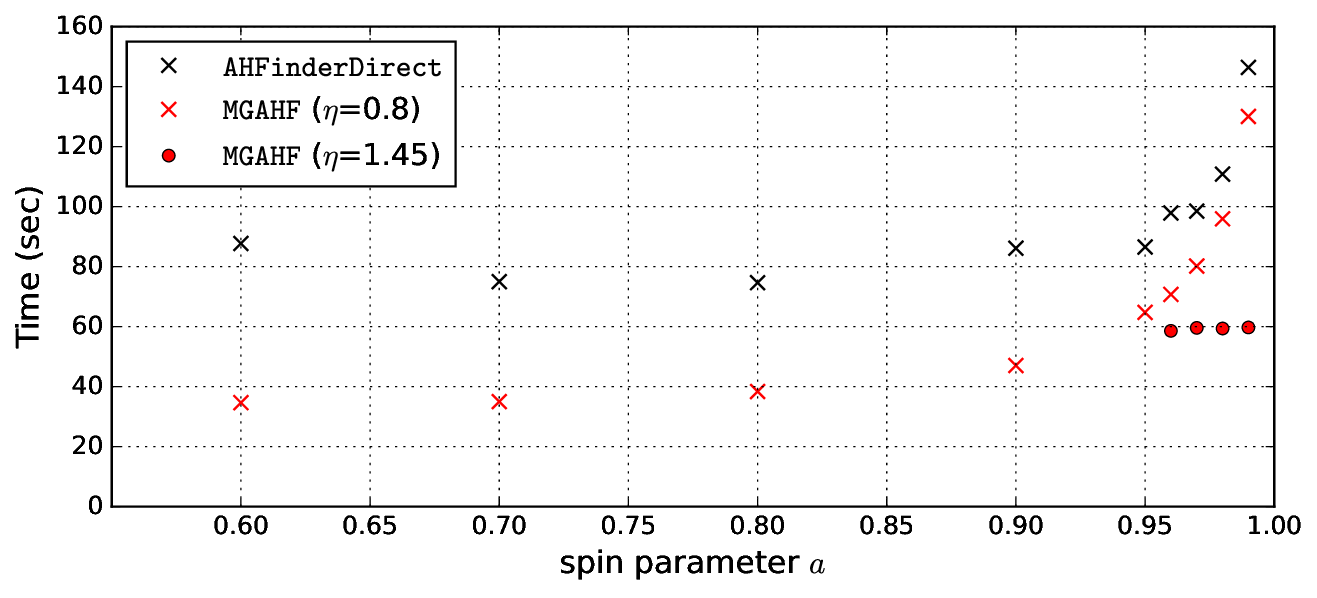}
                \caption{Comparison between our finder and \ahfd\ on the runtime to locate the apparent horizon in numerical off-centered Kerr-Schild data with different values of spin parameter $a$. A spherical surface of radius 1.6 centered at the coordinate origin is used as the initial guess. The angular resolution for the runs is $\SI{1}{\degree}$. The rightmost 4 sets of data points correspond to $a=0.96,0.97,0.98,0.99$.}
                \label{fig:numKS_a99}
            \end{figure}

            At high resolutions, however, we find that our finder generally locates the apparent horizon faster than \ahfd. This is somewhat expected because, at a rough estimation, the operation cost of 
            performing the matrix LU decomposition of the Jacobian equation in \ahfd\ is $\mathcal{O}(N^3)$ but that of one multigrid cycle is $\mathcal{O}(N)$, where $N$ is the total number of grid points used to describe the horizon surface and $N{=}$\tightres{N_\theta}{N_\phi} in our case. The total search time is a competition between the two factors mentioned above. The computational cost of each iteration becomes more important with increasing resolutions. Although the overall convergence rate of our multigrid algorithm is not as fast as Newton's method, there is a critical resolution at which our multigrid algorithm starts to work faster\footnote{%
                \hlRone{There is a variant multigrid algorithm, namely the Newton-multigrid method {\cite{briggs2000}}, that deals with nonlinear equations directly. This method uses an outer Newton step as in AHFinderDirect, but solves the inner Jacobi equation with linear multigrid methods instead. We expect that this should (theoretically) outperform both the algorithms used in AHFinderDirect and our work. We have not used the Newton-multigrid here not only because it requires the manipulation of the Jacobian matrix, but also that it is not directly compatible with the compact finite difference scheme.
                }
            }. 
            In particular, our finder can be more than two times faster when the 
            angular separation between two neighboring grids is about \SI{1}{\degree} as shown in Table \ref{tab:numKS_a0.6}. 
            We have performed the same test for additional values of the spin parameter $a$ at \SI{1}{\degree} resolution, and the runtimes are plotted in Figure \ref{fig:numKS_a99}. Although the runtime of our finder grows with $a$, it takes lesser time to converge in all cases. Recall that the performance of our finder has extra dependence on the convergence parameter $\eta$. At the extreme case of $a=0.99$, the optimal value of $\eta$ shifts to around 1.45, and the runtime is reduced by half comparing to the case $\eta=0.8$.

            We further test on the case where the spin parameter is $a=0.99$, but with the black hole relocated to $(0.2,0.2,0)$. The initial guesses are spherical surfaces centered at the coordinate origin as before, with radius $r_0$ ranging from 1.5 to 1.6 with increments of 0.01.  
            We found that \ahfd\ is more volatile to the initial guess surface, and is able to find the apparent horizon 
            only at $r_0=1.54, 1.57, 1.58$ and $1.6$ (it fails to converge otherwise). On the other hand, our finder is able to find the correct solution for all these cases. 

        \subsubsection{Dynamical binary black hole spacetime}
            We simulate the merger of an unequal-mass black hole binary
            using \textsc{Einstein Toolkit} \cite{ET_paper}. The initial mass ratio of the two non-spinning black holes is set to be $q\simeq 0.7$. We use standard evolution schemes and gauge conditions, with multiple layers of mesh refinements to perform the simulation. 
            The finest grids of grid spacing 
            $\Delta x= \Delta y= \Delta z=0.03125$ cover the two black holes at all times. We export the numerical data to a uniform grid of the finest grid spacing by \textsc{Kuibit} \cite{kuibit} to search for the horizons using our finder. For the two individual horizons, the local coordinate origin is chosen by the position of singularity provided by the \textsc{PunctureTracker} thorn in the simulation code, while that of the common apparent horizon is simply set to be the origin of the simulation grid. 

            \begin{table}[t]
            \caption{\label{tab:numBBH}
            Comparison between our finder and \ahfd\ on the proper areas of the individual horizons and the common apparent horizon found in merging binary black hole spacetime. The relative difference is reported in the last column.}
            \begin{indented}
            \item[]\begin{tabular}{@{}cccc}
            \br
                 & \textsc{MGAHF} & \ahfd & $|\delta A/A|$\\
            \mr
                Horizon of $m_1$ & 
                    19.0239 &
                    19.0097 &
                    \tightnum{7e-4} \\
                Horizon of $m_2$ &
                    \enspace 9.4808 &
                    \enspace 9.4671 &
                    \tightnum{1e-3} \\
                Common apparent horizon & 
                    39.8666 &
                    39.8691 &
                    \tightnum{6e-5} \\
            \br
            \end{tabular}
            \end{indented}
            \end{table}

            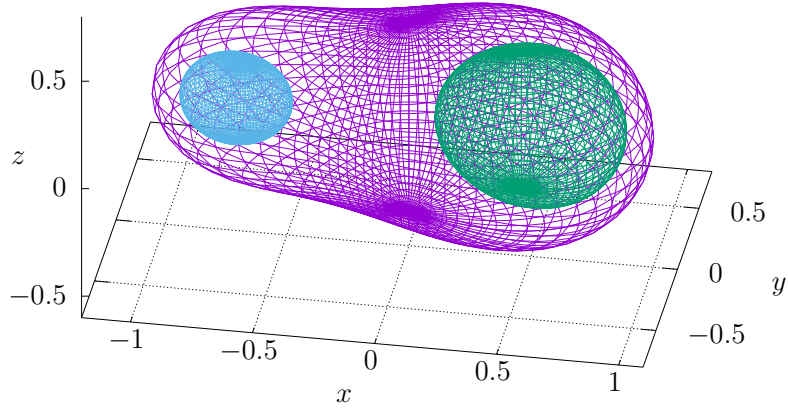
\begin{figure}[t]
                \centering
                \scalebox{.9}{
                  \gdef\gplbacktext{}%
                  \gdef\gplfronttext{}%
                  \ifGPblacktext
                    \def\colorrgb#1{}%
                    \def\colorgray#1{}%
                  \else
                    \ifGPcolor
                      \def\colorrgb#1{\color[rgb]{#1}}%
                      \def\colorgray#1{\color[gray]{#1}}%
                      \expandafter\def\csname LTw\endcsname{\color{white}}%
                      \expandafter\def\csname LTb\endcsname{\color{black}}%
                      \expandafter\def\csname LTa\endcsname{\color{black}}%
                      \expandafter\def\csname LT0\endcsname{\color[rgb]{1,0,0}}%
                      \expandafter\def\csname LT1\endcsname{\color[rgb]{0,1,0}}%
                      \expandafter\def\csname LT2\endcsname{\color[rgb]{0,0,1}}%
                      \expandafter\def\csname LT3\endcsname{\color[rgb]{1,0,1}}%
                      \expandafter\def\csname LT4\endcsname{\color[rgb]{0,1,1}}%
                      \expandafter\def\csname LT5\endcsname{\color[rgb]{1,1,0}}%
                      \expandafter\def\csname LT6\endcsname{\color[rgb]{0,0,0}}%
                      \expandafter\def\csname LT7\endcsname{\color[rgb]{1,0.3,0}}%
                      \expandafter\def\csname LT8\endcsname{\color[rgb]{0.5,0.5,0.5}}%
                    \else
                      \def\colorrgb#1{\color{black}}%
                      \def\colorgray#1{\color[gray]{#1}}%
                      \expandafter\def\csname LTw\endcsname{\color{white}}%
                      \expandafter\def\csname LTb\endcsname{\color{black}}%
                      \expandafter\def\csname LTa\endcsname{\color{black}}%
                      \expandafter\def\csname LT0\endcsname{\color{black}}%
                      \expandafter\def\csname LT1\endcsname{\color{black}}%
                      \expandafter\def\csname LT2\endcsname{\color{black}}%
                      \expandafter\def\csname LT3\endcsname{\color{black}}%
                      \expandafter\def\csname LT4\endcsname{\color{black}}%
                      \expandafter\def\csname LT5\endcsname{\color{black}}%
                      \expandafter\def\csname LT6\endcsname{\color{black}}%
                      \expandafter\def\csname LT7\endcsname{\color{black}}%
                      \expandafter\def\csname LT8\endcsname{\color{black}}%
                    \fi
                  \fi
                    \setlength{\unitlength}{0.0500bp}%
                    \ifx\gptboxheight\undefined%
                      \newlength{\gptboxheight}%
                      \newlength{\gptboxwidth}%
                      \newsavebox{\gptboxtext}%
                    \fi%
                    \setlength{\fboxrule}{0.5pt}%
                    \setlength{\fboxsep}{1pt}%
                \begin{picture}(7200.00,4040.00)
                    \gplgaddtomacro\gplbacktext{%
                      \csname LTb\endcsname
                      \put(1349,584){\makebox(0,0){\strut{}$-1$}}%
                      \csname LTb\endcsname
                      \put(2366,494){\makebox(0,0){\strut{}$-0.5$}}%
                      \csname LTb\endcsname
                      \put(3383,404){\makebox(0,0){\strut{}$0$}}%
                      \csname LTb\endcsname
                      \put(4400,315){\makebox(0,0){\strut{}$0.5$}}%
                      \csname LTb\endcsname
                      \put(5417,225){\makebox(0,0){\strut{}$1$}}%
                      \csname LTb\endcsname
                      \put(6020,625){\makebox(0,0)[l]{\strut{}$-0.5$}}%
                      \csname LTb\endcsname
                      \put(6199,1134){\makebox(0,0)[l]{\strut{}$0$}}%
                      \csname LTb\endcsname
                      \put(6379,1642){\makebox(0,0)[l]{\strut{}$0.5$}}%
                      \put(848,949){\makebox(0,0)[r]{\strut{}$-0.5$}}%
                      \put(848,1844){\makebox(0,0)[r]{\strut{}$0$}}%
                      \put(848,2738){\makebox(0,0)[r]{\strut{}$0.5$}}%
                    }%
                    \gplgaddtomacro\gplfronttext{%
                      \csname LTb\endcsname
                      \put(3161,133){\makebox(0,0){\strut{}$x$}}%
                      \put(6785,1062){\makebox(0,0){\strut{}$y$}}%
                      \put(446,2112){\makebox(0,0){\strut{}$z$}}%
                    }%
                    \gplbacktext
                    \put(0,0){\includegraphics[trim={0 0 0 0},clip]{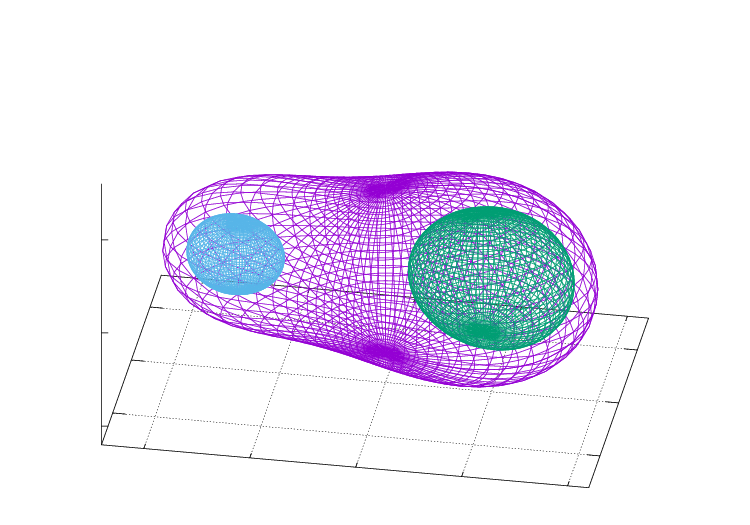}}%
                    \gplfronttext
                  \end{picture}%
                }
                \caption{The common apparent horizon and the individual horizons found by our finder in a dynamical spacetime of merging unequal-mass black hole binary at a time when the common apparent
                horizon first appears.}
                \label{fig:BBH}
            \end{figure}
            
            We report the results of horizon finding at the simulation time when the common apparent horizon first appears in Table \ref{tab:numBBH} and the corresponding snapshot of the three horizon surfaces in Figure \ref{fig:BBH}. The horizon surfaces are modeled by \SI{5}{\degree} angular resolution 
            ($N_\theta{\times}N_\phi = 37{\times}73$). In Figure~\ref{fig:BBH}, the common apparent
            horizon is represented by the purple surface, inside of which are the two individual horizons represented by the green and blue surfaces.                      
            In terms of the proper areas, our finder agrees with the default \ahfd\ in the simulation code to within $0.2\%$. The difference is the biggest for the horizon of the smaller black hole, as the spherical grids become overly stacked at that radius.

\section{Conclusions}
\label{sec:4-conclusion}
    We have presented a new apparent horizon searching algorithm using a multigrid method in this work. We have evaluated the performance of our finder on analytic spacetimes, including the Brill-Lindquist data with two or three black holes. We have recovered the results of the critical parameter for a common apparent horizon to appear in these spacetimes. We have also tested our finder on numerical spacetimes, namely the off-centered Kerr-Schild data and the inspiraling binary black hole spacetime. We have demonstrated that, in terms of speed, the multigrid searching algorithm begins to outperform the currently fastest algorithm (Thornburg's \ahfd\ \cite{Thornburg1996,Thornburg2003}) at high resolutions. We have also shown that our finder is capable of capturing the first appearance of the common apparent horizon in merging black hole binary simulation.
    
    From a numerical perspective of viewing our finder as a general multigrid solver for Poisson equations \hlRone{on the 2-sphere}, there are some notable aspects. We have verified that the line relaxation scheme is preferable (in terms of convergence) to a pointwise relaxation scheme, even when dealing with a solution-dependent nonlinear source term. 
    We have derived the (semi-)compact 4th order finite difference scheme for solving the spherical Poisson equation. The derivation follows a similar 
    procedure used for Cartesian grids (see \cite{HOC_Deriaz2020} and references therein). 
    Our study demonstrates the applicability of this scheme in reducing the complexity of a 
    line relaxation technique on spherical coordinates, from a penta-diagonal matrix inversion to a more efficient tri-diagonal matrix inversion. 
    We believe the formula derived should be directly applicable to other 2-dimensional Poisson equations on the sphere.
    
    We make some final remarks on how our multigrid algorithm can be further improved. 
    First, the segment relaxation scheme (see Section \ref{sec:2C-compactFD}) could be used to improve the convergence. 
    We only use $\phi$-line relaxation as the smoother in our multigrid solver, but the line relaxation direction is preferred to be aligned with the anisotropy direction in the differential operator as it could give better convergence \cite{Barros1991}. 
    Second, using a variant multigrid algorithm might also improve the convergence. We solve the non-linear apparent horizon equation by turning it into a linear elliptic equation with a non-linear source term. In this way, the nonlinearity of the principle equation is avoided.
    However, by not addressing it directly, the rate of convergence could be compromised. 
    The common multigrid methods to tackle nonlinearity directly are the full approximation scheme
    \cite{brandt1977} or Newton-multigrid \cite{briggs2000} algorithm. We have not adopted these methods in the current work since they in principle bring up technical implementation issues in the case of the apparent horizon equation. Exploring these alternative multigrid schemes could be one future investigation direction for increasing the efficiency of our multigrid code.

\ifshowall{
    \section*{Data availability statement}
    No new data were created or analyzed in this study.
    
    \section*{Acknowledgments}
    We thank Kenneth Chen and Yun-Kau Lau for useful discussions. LML is also grateful for the hospitality of the Morningside Center of Mathematics, Chinese Academy of Sciences, where the idea of this project was initiated. 
    This work is partially supported by a grant from the Research Grants Council of the Hong Kong SAR, China [Project No.~CUHK 14304322]
}\fi

\section*{ORCID iDs}
Hon-Ka Hui\,\orcidlink{0000-0002-2815-527X} \url{https://orcid.org/0000-0002-2815-527X}\\
Lap-Ming Lin\,\orcidlink{0000-0002-4638-5044} \url{https://orcid.org/0000-0002-4638-5044}

\section*{References}
\bibliography{References}
\ifpreprint\stepcounter{footnote}\footnotetext{This preprint is prepared with the IOP Publishing \LaTeXe\ preprint class.}\fi

\newpage
\appendix
\section{\textit{Fully}-compact 4th-order finite difference scheme for spherical Poisson equations}
\label{appendix:compactFD}
    Suppose we solve a general Poisson equation in spherical coordinates,
    \begin{equation}
    \label{app_eq:Poisson}
        \left( \pdv[2]{}{\theta} + \cot\theta \pdv{}{\theta} + \frac{1}{\sin^2\theta} \pdv[2]{}{\phi} \right) u(\theta, \phi) = f(\theta, \phi),
    \end{equation}
    on a spherical grid of grid size\footnote{%
        We follow the standard notation to use $h$ for the grid spacing, which should not be confused with the horizon function $h(\theta, \phi)$ used elsewhere in the paper.} 
    $h=\Delta\theta$ and $k=\Delta\phi$. Using a 2nd-order finite difference approximation, the first two terms on the left side are
    \begin{equation}
    \label{app_eq:compactFD_theta_deriv}
        \fl
        \pdv[2]{u}{\theta} + \cot\theta \pdv{u}{\theta}
        = \left( \pdv[2]{u}{\theta} + \cot\theta \pdv{u}{\theta}\right)_{\rm FD2}
        - \frac{h^2}{12} \left( \pdv[4]{u}{\theta} + 2\cot\theta \pdv[3]{u}{\theta} \right) + \mathcal{O}(h^4).
    \end{equation}
    We then seek an expression to replace the third and fourth order derivatives in the second parenthesis. Differentiating (\ref{app_eq:Poisson}) w.r.t $\theta$ gives
    \begin{equation}
        \pdv[3]{u}{\theta} + \cot\theta \pdv[2]{u}{\theta} - \frac{1}{\sin^2\theta} \pdv{u}{\theta}
        -\frac{2\cos\theta}{\sin^3\theta} \pdv[2]{u}{\phi} + \frac{1}{\sin^2\theta} \frac{\partial^3u}{\partial\theta\partial\phi^2}
        = \pdv{f}{\theta}.
    \end{equation}
    The second derivative reads
    \begin{eqnarray}
        \pdv[4]{u}{\theta} &+ \cot\theta \pdv[3]{u}{\theta} - \frac{2}{\sin^2\theta} \pdv[2]{u}{\theta} + \frac{2\cos\theta}{\sin^3\theta} \pdv{u}{\theta} \nonumber \\
        &+ \frac{6-4
        \sin^2\theta}{\sin^4\theta}\pdv[2]{u}{\phi}
        -\frac{4\cos\theta}{\sin^3\theta} \frac{\partial^3u}{\partial\theta\partial\phi^2}
        + \frac{1}{\sin^2\theta} \frac{\partial^4u}{\partial\theta^2\partial\phi^2}
        = \pdv[2]{f}{\theta}.
    \end{eqnarray}
    Combining the last two equations, we have
    \begin{eqnarray}
        \fl
        \pdv[4]{u}{\theta} + 2\cot\theta \pdv[3]{u}{\theta} 
        ={}& \left( \cot\theta \pdv{f}{\theta} + \pdv[2]{f}{\theta} \right)
        + \left( 1+\frac{1}{\sin^2\theta} \right) \pdv[2]{u}{\theta} 
        - \frac{\cos\theta}{\sin^3\theta} \pdv{u}{\theta}
        \nonumber \\
        &+ \frac{2\sin^2\theta-4}{\sin^4\theta} \pdv[2]{u}{\phi} 
        + \frac{3\cos\theta}{\sin^3\theta} \frac{\partial^3u}{\partial\theta\partial\phi^2} 
        - \frac{1}{\sin^2\theta} \frac{\partial^4u}{\partial\theta^2\partial\phi^2}.
    \end{eqnarray}
    The right hand side of this equation can be evaluated by 2nd order finite difference approximation and be substituted back to (\ref{app_eq:compactFD_theta_deriv}) while the overall error is maintained at $\mathcal{O}(h^4)$. We can then combine the result with (\ref{eq:compactFD_phi_deriv}) to obtain the \textit{fully}-compact 4th order finite difference approximation for the spherical Poisson equation (\ref{app_eq:Poisson}). The explicit expression is
    \begin{eqnarray}
        \fl
        \left( \pdv[2]{u}{\theta} + \cot\theta \pdv{u}{\theta} \right)_{\rm FD2} + \left( \frac{1}{\sin^2\theta} \pdv[2]{u}{\phi} \right)_{\rm FD2}
        -\frac{k^2}{12} \left[ \pdv[2]{f}{\phi} - \frac{\partial^4u}{\partial\theta^2\partial\phi^2} - \cot\theta \frac{\partial^3u}{\partial\theta\partial\phi^2} \right]_{\rm FD2}
        \nonumber \\
        \fl
        -\frac{h^2}{12} \Bigg[
            \cot\theta \pdv{f}{\theta} + \pdv[2]{f}{\theta} 
            + \left( 1+\frac{1}{\sin^2\theta} \right) \pdv[2]{u}{\theta}
            - \frac{\cos\theta}{\sin^3\theta} \pdv{u}{\theta}
            + \frac{2\sin^2\theta-4}{\sin^4\theta} \pdv[2]{u}{\phi} 
            \nonumber \\
            + \frac{3\cos\theta}{\sin^3\theta} \frac{\partial^3u}{\partial\theta\partial\phi^2} 
            - \frac{1}{\sin^2\theta} \frac{\partial^4u}{\partial\theta^2\partial\phi^2}
        \Bigg]_{\rm FD2}
        + \mathcal{O}(h^4, h^2k^2, k^4) \quad = f.
    \end{eqnarray}


\end{document}